\documentclass[aps,amsfonts,nofootinbib,superscriptaddress,preprint]{revtex4-1}
\usepackage[normalem]{ulem} %% Para tachar, comando sout \sout{for the first kinematics}
\makeatletter
\usepackage{amsmath}
\usepackage{cleveref}
\usepackage{amssymb}
\usepackage{amsbsy}
\usepackage{bm}
\usepackage{epsfig} 
\usepackage{color}
\usepackage{slashed}
\usepackage{soul}
\usepackage{comment}
\makeatletter
\usepackage{appendix}
\usepackage{epsfig} 

\newcommand{\stkout}[1]{\ifmmode\text{\sout{\ensuremath{#1}}}\else\sout{#1}\fi}

\newcommand{\ee}{\end{equation}}
\newcommand{\bb}{\begin{equation}}
\newcommand{\eqb}{\begin{eqnarray}}

\newcommand{\eqf}{\end{eqnarray}}

\usepackage[T1]{fontenc}

\begin{document}
\title{  Spontaneous Symmetry Breaking and the Cosmological Constant}

\author{J. Gamboa
}
\email{jorge.gamboa@usach.cl}
\affiliation{Departamento de F\'isica, Universidad de Santiago de Chile, Casilla 307, Santiago, Chile}
\author{J. Lopez-Sarri\'on}
\email{justo.lopezsarrion@ub.edu}
\affiliation{Departament de Física Quàntica i Astrofísica and
Institut de Ciències del Cosmos (ICCUB), Universitat de Barcelona,
Martí Franquès 1, 08028 Barcelona, Spain}
\author{F. M\'endez}
\email{fernando.mendez@usach.cl}
\affiliation{Departamento de F\'isica, Universidad de Santiago de Chile, Casilla 307, Santiago, Chile}
\author{N. Tapia-Arellano}
\email{u6044292@utah.edu}
\affiliation{Department of Physics and Astronomy, University of Utah, Salt Lake City, UT 84102, USA}
\begin{abstract}
An approach that allows studying the relationship between the neutralization of the cosmological constant and instantons for cosmology coupled to antisymmetric fields  is proposed. Using suitable variables, the Lagrangian leading to the FRW equations can be written analogously to a Ginzburg-Landau theory and we show how spontaneous symmetry breaking appears. The approach leads to three possible solutions, if we denote $\Lambda_1$ and 
$\Lambda_2$ as the cosmological constants that come from gravity ($\Lambda_1$) and antisymmetric fields ($\Lambda_2$), then the solutions, a) $ |\Lambda_2|>|\Lambda_1|$ is consistent with a very small but non-zero $\Lambda_1$, b) $|\Lambda_2|<|\Lambda_1|$, then $\Lambda_1$  is observationally ruled out and, c) $|\Lambda_2|=|\Lambda_1|$ is a solution with finite action and instantons restoring the symmetry $\varphi \to -\varphi$. The third solution is also consistent with neutralization of the cosmological constant in four dimensions.
\end{abstract}
\maketitle

%%%%%%%%%%%%%%%%%%%%%%%
The problem of the cosmological constant is one of the crucial challenges of modern physics. Its solution, or even a different approach to this problem, can help to understand other cosmological puzzles that until now have not been successfully tackled \cite{weinberg,carroll,coleman1}. For over  forty years, it has been known that the coupling of antisymmetric fields to Einstein's theory induces an effective cosmological constant \cite{nicolai,duff,freund} which might even be neutralized due to the presence of 
such fields.

The effective cosmological constant, $\Lambda_{eff}$, must be positive for the universe to  accelerate (for a review, see \cite{abdalla}). It is composed of different contributions due to the presence of the antisymmetric fields, and such components have opposite signs  to make the neutralization occurs. In four dimensions, there are no global obstructions \cite{nicolai,duff,freund}, and the kinetic term of these fields behaves as  $\sqrt{g}\,F_{\mu \nu \lambda \rho}F^{\mu \nu \lambda \rho} \sim  \Lambda_2\, \sqrt{g}$, {\it i.e.}, producing  the effective cosmological constant 
$\Lambda_{\text eff}=\Lambda_1+\Lambda_2$.
%The neutralization to occur if one of the parts of the effective cosmological constant %has an opposite sign and --at the same time-- satisfies 
%$\Lambda_{{\text eff}}>0$ for the universe to be accelerating (for a review see 
%\cite{abdalla}). In four dimensions there are no global problems 
%\cite{nicolai,duff,freund} and the kinetic energy $\sqrt{-g} F_{\mu \nu \lambda 
%\rho}F^{\mu \nu \lambda \rho} \sim  \Lambda_2 \sqrt{-g}$, {\it i.e.} produces the 
%effective cosmological constant $\Lambda_{\text eff}=\Lambda_1+\Lambda_2$.

In higher dimensions the problem is more subtle because the $4$-form $F$ satisfies a generalized Dirac quantization condition \cite{,Teitelboim,Bousso} and the cosmological constant neutralization problem requires a discrete parameter space. 

A first mechanism implementing such ideas has been  discussed in  \cite{nicolai,duff,freund}  for a  four-dimensional spacetime. In contrast, a second one,
analyzed in  \cite{ct1}, is defined in  higher dimensions containing  some of the essential ingredients introduced in \cite{Bousso}. For other approaches, see for example \cite{kaloper0,kaloper00,kaloper,otros}  and references therein. 

In this paper, we want to examine the problem of neutralizing the cosmological constant 
using conventional physics.   What does conventional physics mean?  It means, 
surprisingly, that by writing suitably $-\Lambda_2 \sqrt{-g}$ (the cosmological term coming 
from $F_{\mu \nu \rho \lambda}F^{\mu \nu \rho \lambda}$ in four dimensions) the 
cosmological model  fits precisely  with a Ginzburg-Landau theory, the most prominent 
description of an effective classical field theory.

%The above assertions are nicely exemplified using some standard cosmology results because 
%--as a consequence of the cosmological principle-- the equations are considerably 
%simplified and  reinterpreted more directly.    

In order to explain our results, we begin by considering the Euclidean action of Einstein's theory 
% In this note we will investigate a cosmological constant reduction mechanism for the FRW universe coupled to relativistic membranes (actually $D$-branes in a $D+1$-dimensional spacetime) as a way to analyze spontaneous symmetry breaking in cosmology. and the role played by instantons.
%
%In order to explain our starting point we note that the $D$-brane has the form $T \int d^Dx\, \sqrt{g}$, this makes the cosmological constant writeable as an effective constant because plugging in the euclidean Einstein action we get
   \bb 
   S= \int d^Dx \, \sqrt{-g}\left( R-2 \Lambda_{eff} \right) + \cdots . \label{1}
   \ee 
   where $\Lambda_{eff}= \Lambda_1+\Lambda_2$ and $8\pi G=1$. In the previous 
   expression,  the term \lq\lq $\cdots$ \rq\rq\, denotes other  possible couplings.
   
   Under these conditions, we will show that the cosmological constant can be dynamically adjusted and eventually neutralized.
   
 Let us consider the FRW universe with zero spatial curvature 
   \bb
   ds^2 =- N^2(t) dt^2 + a^2(t) \left(dx^2 + dy^2 +dz^2 \right), \label{2}
   \ee
with  $a(t)$ the scale factor and $N(t)$, an auxiliary field that guarantees  the 
invariance of the action under time reparametrization.

The Lagrangian leading to the FRW equations \cite{gondolo}
   \eqb 
   &&2 \frac{\ddot a}{a} + \left(\frac{{\dot a}}{a}\right)^2 + \Lambda_{eff} = 0, \label{3}
   \\
   && \left(\frac{\dot a}{a}\right)^2   =\frac{\Lambda_{eff}}{3}, 
   \label{4}
   \eqf
   is
   \bb
   L= \frac{a {\dot a}^2}{2 N} + \frac{1}{6} \Lambda_{eff} N a^3. \label{5}
   \ee
   
It is worth noting that the $\Lambda_2$ part contained in $\Lambda_{eff}$ can  be obtained from  a sort of relativistic membrane, with $\Lambda_2$
playing the role of the tension. Indeed, consider the  Lagrangian $L_{\Lambda_2} = \Lambda_2 \sqrt{-g}$, which can also be obtained from   
\footnote{This procedure was used on the gauge $\lambda=1$ in \cite{eguchi} to obtain the functional diffusion equation in string theory (for the extension to $D$-brane see \cite{marti1}. See also \cite{ctm,Lizzi}).}

   \bb
   \tilde{L}_{\Lambda_2}=\frac{1}{2\lambda} g + \frac{1}{2}\Lambda^2_2 \lambda, \label{7}
   \ee
   with $\lambda(t)$ an auxiliary field. The Lagrangian (\ref{7}) has the advantage of having a smooth $\Lambda_2\to 0$ limit and is analogous to the one of the massive relativistic particles in the einbein formulation. Namely $\tilde{L}_\lambda= \frac{1}{2 \lambda} {\dot x}^2 + \frac{1}{2}m^2 \lambda$, instead of     $L_\lambda=-m\sqrt{{-\dot x}^2}$. Here,  $\tilde{L}_\lambda$ has the advantage of having a smooth massless   limit.

   The contribution (\ref{7}) in the FRW geometry, gives rise to  the full Lagrangian
  \bb
   L= \frac{a {\dot a}^2}{2 N} +\frac{1}{6} \Lambda_1 N a^3 + \frac{1}{12\lambda} a^6 N+ \frac{\Lambda^2_2}{12} \lambda N, \label{8}
   \ee
   which, with the change of variables
  \bb \varphi= \frac{2}{3} a^{\frac {3}{2}}, \label{9}
\ee 
turns out to be
   \bb
   L= \frac{1}{2N}{\dot \varphi}^2 + \frac{3}{8} \Lambda_1  N \varphi^2 + \frac{27}{64\lambda}N \varphi^4 + \frac{1}{12} \Lambda_2^2 \lambda N. \label{10}
   \ee
   
   The  potential has two minima,  
    \bb
   \varphi_0=\pm \frac{2 }{3 }\sqrt{-{{\lambda}}\,\Lambda_1 }, \label{11}
   \ee 
 such that  perturbations around these minima  break the 
 symmetry $\varphi \to -\varphi$.
 
 It is convenient to define   $-\mu^2= \frac{3}{4}\Lambda_1$, and
 $\frac{\kappa}{4}= \frac{27}{{64}}$. Then \cref{10} becomes
 \eqb
   L&=& = \frac{1}{2N}{\dot \varphi}^2 - \frac{\mu^2}{2}   N\, \varphi^2 + \frac{\kappa}{4}\, \frac{N}{\lambda}\varphi^4 + \frac{1}{12} \,N\,\lambda\,\Lambda_2^2,  \label{a10}
   \\
   &=& L_0 +  \frac{1}{12} N\,\Lambda_2^2,  \label{aa10}
      \eqf  
leading to the field equation
 \bb
\frac{d}{dt}\left(\frac{\dot \varphi}{N}\right) + \mu^2 N \,\varphi -\frac{\kappa N}{\lambda} \varphi^3=0, \label{eq1}
 \ee 
 and the reparametrization constraint
 \bb
 \frac{{\dot {\varphi}}^2}{{{2}}N^2} + \frac{\mu^2}{2} \varphi^2 - \frac{\kappa}{4 } \varphi^4 
 -
 \frac{\lambda}{12} \Lambda_2^2=0. \label{a2}
 \ee
 
 These equations are  simplified by choosing the `proper-time' gauge $N=1$, and $\lambda=1$ \cite{marti1} to obtain
 \eqb
 {\ddot \varphi} + \mu^2  \,\varphi -\kappa \varphi^3&=&0, \label{eq111}
 \\
\frac{1}{2}{\dot {\varphi}}^2 + \frac{\mu^2}{2} \varphi^2 - \frac{\kappa}{4 } \varphi^4 - \frac{1}{12} \Lambda_2^2&=&0. \label{a22}
 \eqf 
  
 The first equation has the solution
   \bb
   \varphi = \frac{\mu}{\sqrt{\kappa}} \tanh \left[ \frac{\mu}{\sqrt{2}} (t-t_0)\right]. \label{rela5}
   \ee

The solutions of the  constraint (\ref{a22}), on the other hand, are solutions
of (\ref{eq111}). Indeed, multiplying this equation by $\dot{\varphi}$,
gives rise to 
\[
\frac{d}{dt} \left[\frac{1}{2}{\dot {\varphi}}^2 + \frac{\mu^2}{2} \varphi^2 - \frac{\kappa}{4 } \varphi^4 \right]=0,
\]
and therefore, $\frac{1}{2}{\dot {\varphi}}^2 + \frac{\mu^2}{2} \varphi^2 - \frac{\kappa}{4 } \varphi^4 $ is a constant which turns out to be $\mu^4/4\kappa$, and must be equal to  $\frac{1}{12}\Lambda_2^2$,  imposing  the condition $|\Lambda_1|=|\Lambda_2|$, namely the neutralization of the effective cosmological constant, as we will discuss in what follows. 
      
      In Euclidean space, the Lagrangian is the Gibbs free energy (density) and (\ref{aa10}) can be viewed in the proper-time gauge as a Ginzburg-Landau theory with the parameters $\mu^2$ and $\kappa$ formally as functions of $T$.  
 
Global stability demands that $\kappa>0$ while $\mu^2$, according the Ginzburg-Landau theory, becomes 
 $
 \mu^2 \to \mu^2(T)= \mu^2_0 \left(1-\frac{T}{T_c}\right)  \label{b10}
 $,
 where $\mu_0^2$ is some function of $T$ although its sign is fixed by the cosmological constant \footnote{We do not discuss this issue in depth here, but the study of the phase transition including sources and inflation is an aspect that we hope to discuss in a future paper.}. 
 
 In (\ref{aa10}), $L_0$ is conventional Gibbs free energy while $\frac{1}{12}\Lambda_2^2$  is analogous to an external field and it does not play any role for the phase transition. 
 
 If we take this fact into account, we can choose the combination of parameters $\{\mu^2,\Lambda_2\}$ such that 
    \bb
  L= \frac{1}{2}{\dot \varphi}^2 + \frac{\kappa}{4}\left(\varphi^2-\frac{\mu^2}
  {{{\kappa}}} \right)^2 + \Delta L[\mu,\Lambda_2,T], \label{14} 
  \ee  
  where 
  \eqb
 \Delta L&=& -\frac{\mu^4}{4 \kappa}+ \frac{1}{12}\Lambda_2^2 \nonumber
 \\
 &=& \frac{1}{12}(\Lambda_2^2 -\Lambda_1^2). \label{rela1}
  \eqf
  %The equation (\ref{rela1}) 
  giving  rise to the following possibilities 
  %can be summarized as follows:  
 
 \begin{itemize}
 \item  $\Delta L\geq 0$ implies $|\Lambda_2| \geq|\Lambda_1| $, therefore, if the cosmological constant $\Lambda_2$ is very small, then $\Lambda_1$ that comes from the gravity sector should be even smaller (although not zero).
 \item If  $\Delta L\leq 0$ then $|\Lambda_2| \leq|\Lambda_1| $. This possibility however produces a very large cosmological constant  ($\Lambda_1$)
 and is ruled out observationally. 
 \item   Finally,  $\Delta L=0$  results 
 \bb 
 \Lambda_1 =  \Lambda_2. \label{condi20}
 \ee  
 
 This equation has the consequence that the action $\int_{-\infty}^{\infty} dtL$ is finite and therefore implies  the existence of instantons \cite{Polyakov,Coleman}.  It is the instantons that restore this symmetry breaking $\varphi \to -\varphi$. 
  \end{itemize}

\begin{figure}
\begin{center}
\includegraphics[scale=0.5]{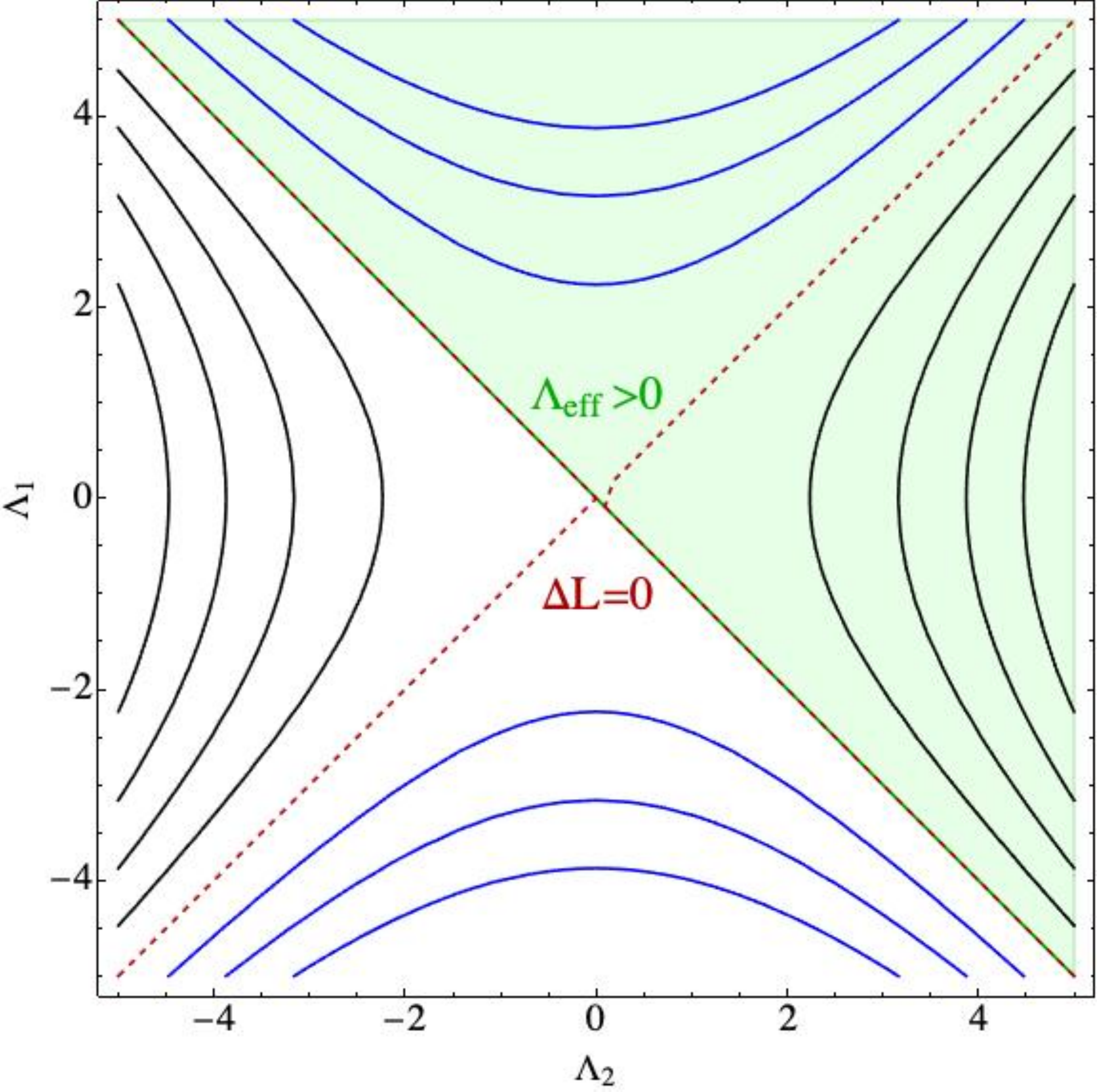}
\caption{\small{The figure shows a possible qualitative plot  of the effective cosmological constant for the three scenarios given by (\ref{rela1}). The possibilities show some values of the effective cosmological constant.
The green region shows the physical condition $\Lambda_{eff}=\Lambda_1 + \Lambda_2 > 0 $.
Black, blue and red show conditions for $\Delta L$. Black lines correspond to $\Delta L > 0$, blue shows $\Delta L <0$ and red shows $\Delta L = 0$, in the neutralization scenario.}}
\label{fig:regions}
\end{center}
\end{figure}

  In terms of $\Lambda_2$ the effective cosmological constant has the following solution   
   \bb
 -2|\Lambda_2| \leq \Lambda_{\text{eff}} \leq  2|\Lambda_2|. \label{ine1}
   \ee
  which shows that there are regions between $-2|\Lambda_2|$ and $2|\Lambda_2|$ where $\Lambda_{\text eff}$ is neutralized (see \Cref{fig:regions}).  

 Summarizing, we highlight that if (\ref{condi20}) holds, the instantons do the dirty work of restoring the symmetry breaking and  provide a novel approach to the question of the neutralization of the cosmological constant.

     \section*{Acknowledgements}
We would like to thank A. P. Balachandran, J. Jaeckel, N. Kaloper and V. P. Nair for the questions and suggestions. This research was supported by DICYT 042131GR and Fondecyt 1221463 (J.G.), 042231MF (F.M.), JLS thanks the Spanish Ministery of Universities and the European Union Next Generation EU/PRTR for the funds through the Maria Zambrano grant to attract international talent 2021 program.
The work of NTA is supported by the University of Utah.

\end{document}